\documentclass{elsart}

\usepackage{epsfig}

\begin{document}


\begin{frontmatter}

%
%
\title{Self-pinching of a relativistic electron bunch \\ in a drift tube}

%
%
\author{Claudio G. Parazzoli and Benjamin E.C. Koltenbah}

%
%
\address{The Boeing Company,
         Boeing Defense \& Space Group,
         P.O. Box 3999, M/S 87-85 \\
         Seattle, WA  98124-2499,
         USA}

%
%
\begin{abstract}
Electron bunches with charge densities $\rho$ of the order of
$10^2$ to $10^3$ [nC/cm$^3$], energies between $20.$ and $100.$ [MeV],
peak current $> 100$ [A], bunch lengths between $0.3$ and $1.8$ [cm],
 and bunch charge of $2.0$ to $20.$ [nC] are relevant to the design of
Free Electron Lasers and future linear colliders.
In this paper we present the results of numerical simulations performed
with a particle in a cell (pic) code of an electron bunch in a drift tube.
The electron bunch has cylindrical symmetry with the $z$-axis oriented 
in the direction of motion.
The charge density distribution is constant in the radial and Gaussian in
the longitudinal direction, respectively.
The electron bunch experiences both a radial pinch in the middle of the
pulse, corresponding to the peak electron density, and a significant growth
of the correlated emittance.
This behavior is explained, and an approximate scaling law is identified.
Comparisons of the results from the pic and PARMELA codes are presented.
\end{abstract}

\end{frontmatter}


%
%

\section{Introduction}
New designs of Free Electron Lasers and high-brightness colliders
require the modeling of electron bunches with higher charge densities than
have previously been studied.
In this high-charge regime, self-fields must be handled consistently.
``Particle-in-a-cell'', or pic, codes must therefore be employed instead of
the commonly-used ``particle-pusher" codes, such as PARMELA, which do not
account completely for self-fields.
We studied the simple case of a bunch traveling along a drift
tube in order to gain a better understanding of the self-field effects in
this high-charge regime.
We have completed a detailed series of calculations using a pic code
(detailed below) on a range of energy and electron charge
where significant disagreement between pic and ``particle-pusher'' codes
has been found.
Our results display rather severe self-pinching
of the bunch under certain circumstances, implying the existence of a
radially inward-directed force which
we explain analytically.

The outline of the paper is as follows:
Section \ref{sec:theory} contains a discussion of the forces acting on an
electron bunch, the origin of the radial pinching, and an estimate of the
initial space charge potential.
Section \ref{sec:numerics} contains numerical results of the pic code, a
discussion of the scaling parameters, and comparison between the pic and
PARMELA codes results.  
Section \ref{sec:summary} contains a summary of the paper and conclusions.

%
%

\section{Radial pinching force and space charge potential}
\label{sec:theory}
\begin{figure}[h]
\begin{center}
\epsfig{file=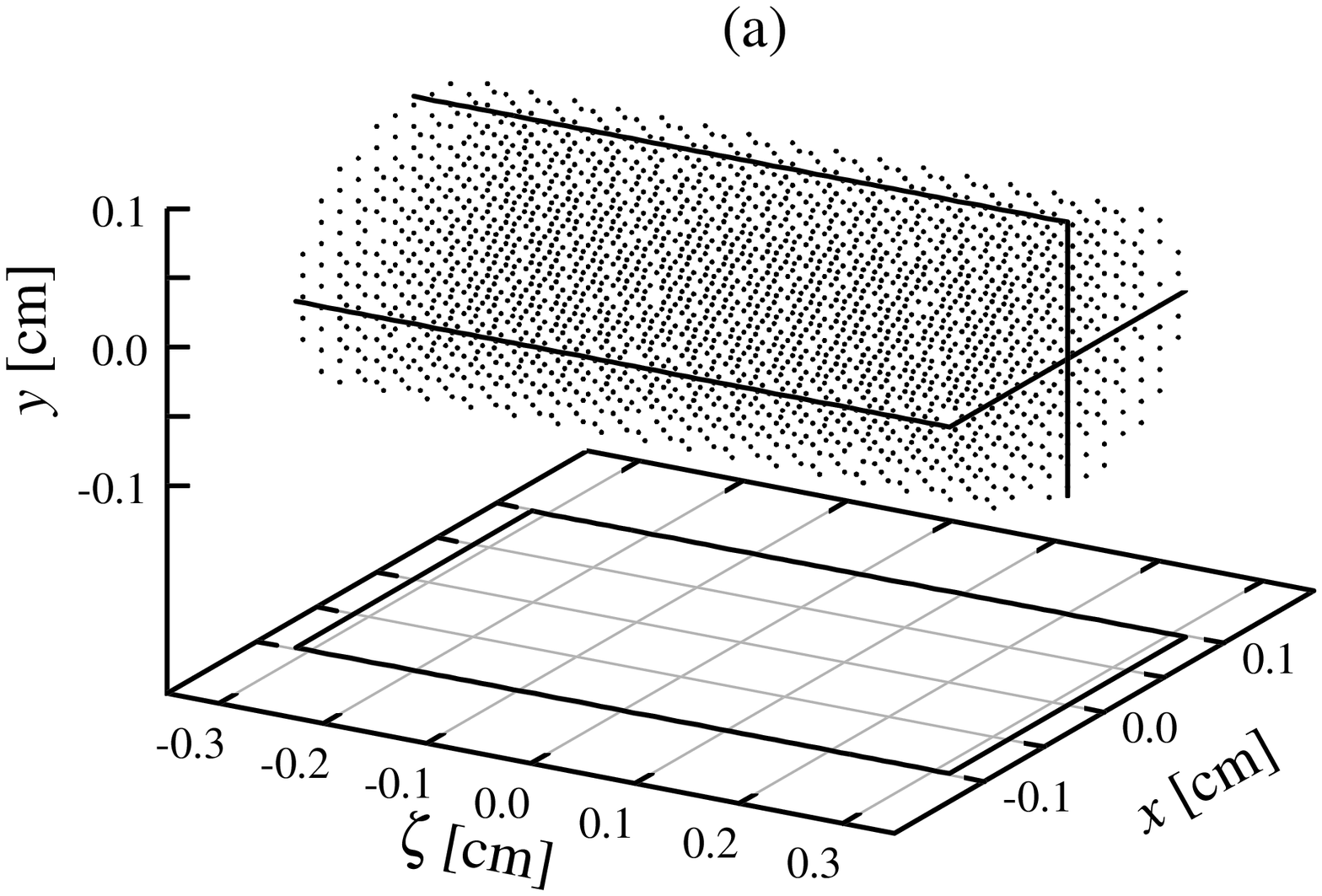,width=2.7in,bb=78 124 637 508,clip=true}
\epsfig{file=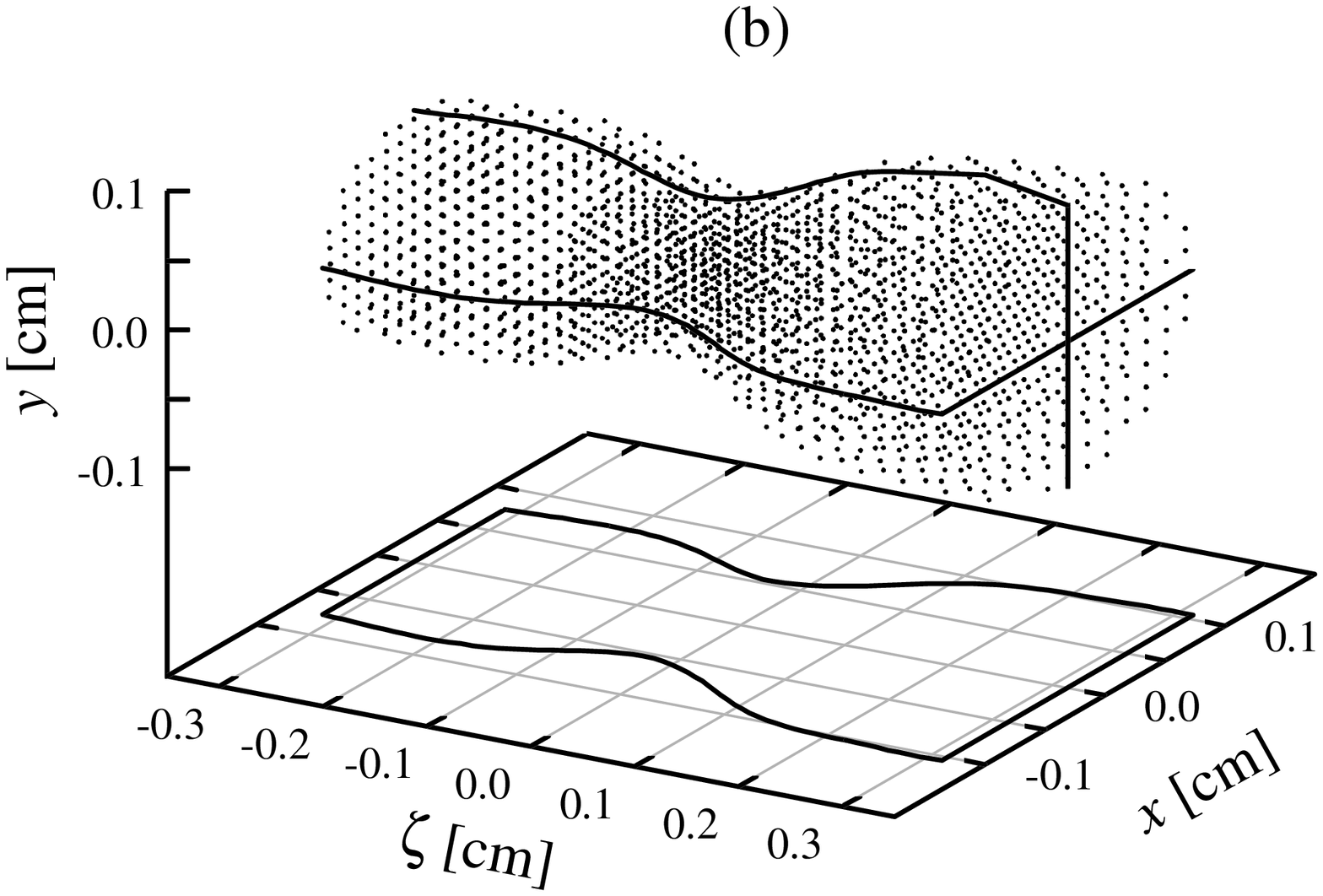,width=2.7in,bb=78 124 637 508,clip=true}
\end{center}
\caption{Electron bunch and bunch envelope at drift positions (a) $z_c = 0$
[cm] (drift entrance) and (b) $z_c = 30.$ [cm] (drift exit).  The pulse
envelope is projected on the bottom panel.  Bunch parameters: $Q = 2.0$
[nC], $a = 0.1$ [cm], $\sigma = 0.1274$ [cm], $b = 1.5$ [cm], $E = 20.0$
[MeV], $\gamma_0 = 40.138$ .}
\label{fig:bunch1}
\end{figure}

It is well known \cite{Tao,Reassure} that for an infinitely long
relativistic electron beam, traveling in the positive $z$-direction and
having constant electron density along $z$, the outward-directed radial
space charge force is nearly balanced by the inward-directed Lorentz force.
The electron beam is thus subjected to a small net defocusing effect.
In this section we show, for the case of a finite length electron bunch,
the origin of a radial force which is directed inward and which generates
a pinch in the electron bunch envelope as seen in Figure \ref{fig:bunch1}.
The motion of the particles is in the positive $z$-direction, 
cylindrical symmetry is assumed, and the relative longitudinal particle 
position in the bunch is denoted by $\zeta$, where $\zeta = z - z_c$, and
$z_c$ is the $z$-position of the center of the bunch.

%
%

We assume, following \cite{Morton}, that all the electrons travel along
$z$ with uniform  velocity $v = \beta_0 c$, no transverse velocity
component is present, and the density $\rho(r,z)$ is given by
\begin{equation}
  \rho(r,s(z,t)) =
  \left\{
    \begin{array}{c@{\quad:\quad}l}
      Q / (\pi a^2)\ f(s) = n_o f(s)  &  r < a \\
      0                               &  r > a,
    \end{array}
    \right.                                              \label{eq:density}
\end{equation}
where $s = z - \beta_0 c t$,
$f(s)= \e^{-(s/\sigma)^2/2} / [(2 \pi)^{1/2} \sigma]$
is the normalized longitudinal density distribution, $Q$ is the pulse
charge with appropriate sign, and $a$ is the electron bunch radius.
We further assume that the $z$- and $t$-dependencies of the electric field
$E$ and magnetic field $B$ in Maxwell's equations are combined in the
variable $s$.
This {\em ansatz} is appropriate when the electrons have a common
longitudinal velocity $\beta_0$ and no transverse motion.
This assumption is violated, as shown by the numerical simulations, after
the bunch drifts a significant distance.
It is, however, useful to establish the conditions for the onset of the
radial pinching force.
The region in which we will solve Maxwell's equations is the inside of a
perfectly conductive drift tube of radius $b$.

In view of the symmetry of the problem, only the $E_r^0$, $E_x^0$, and
$B_\theta^0$ components of the fields are present.
The superscript $()^0$ is used to indicate that the fields are computed in
the absence of any transverse velocity.

The Fourier transform of the electric field is
\begin{equation}
  \vec{\mathcal{E}}(k,r) = (2 \pi)^{-1}  \int \vec{E}(r,s)
    \e^{-\mathrm{i}ks} \d s,
\end{equation}
and similar definitions hold for the Fourier transforms
$\vec{\mathcal{B}}(k,r)$ and $\mathcal{F}(k)$ of $\vec{B}(r,s)$ and $f(s)$,
respectively.
For our selection of $f(s)$, a Gaussian distribution with standard
deviation $\sigma$ is assumed, and $\mathcal{F}(k) = \e^{-(k\sigma)^2/2}$.

Upon substitution into Maxwell's equations, $\mathcal{E}_r^0$ and
$\mathcal{B}_\theta^0$ can be expressed as functions of $\mathcal{E}_z^0$
only, and inside the perfectly conductive drift tube the results are as
follows:
\begin{eqnarray}
  \mathcal{E}_r^0 & = & \frac{\mathrm{i}k}{q^2}
    \frac{\partial \mathcal{E}_z^0}{\partial r},          \label{eq:er0} \\
  \mathcal{B}_\theta^0 & = & \frac{\mathrm{i}kb}{q^2}
    \frac{\partial \mathcal{E}_z^0}{\partial r},         \label{eq:btheta0}
\end{eqnarray}
where $q = -\mathrm{i} k/\gamma_0$ and $\gamma_0^2 = (1 - \beta_0^2)^{-1}$.
In the region inside the electron bunch, where $r < a$, the differential
equation satisfied by $\mathcal{E}_z^0$ is
\begin{equation}
  \frac{1}{r} \frac{\partial}{\partial r}
    \left( r \frac{\partial \mathcal{E}_z^0}{\partial r} \right)
  + q^2 \mathcal{E}_z^0 =
  \mathrm{i} \frac{4 \pi n_0}{\gamma_0^2} k \mathcal{F}(k),
                                                         \label{eq:diffeq1}
\end{equation}
and in the region outside the electron bunch, where $a < r < b$,
\begin{equation}
  \frac{1}{r} \frac{\partial}{\partial r}
    \left( r \frac{\partial \mathcal{E}_z^0}{\partial r} \right)
  + q^2 \mathcal{E}_z^0 = 0.                             \label{eq:diffeq2}
\end{equation}
The solution of Eqs. (\ref{eq:diffeq1}) and (\ref{eq:diffeq2}) must be
finite, continuos everywhere, vanish at the conductive wall of the drift
tube ($r = b$) and insure the continuity of $\mathcal{E}_r^0$ at the edge
of the pulse ($r = a$).
The solution is
\begin{equation}
  \mathcal{E}_z^0(k,r) =
  \left\{
    \begin{array}{l@{\quad:\quad}l}
      C_1 J_0(qr) + 4\pi n_0 \ \mathcal{F}(k)/(\mathrm{i} k) &
      r < a \\
      \left. \begin{array}{l}
        [C_1 - 2\pi^2 n_0 q a Y_1(q a) \ \mathcal{F}(k) / (\mathrm{i} k)]
          \ J_0(q r) \\
        + [2\pi^2 n_0 q a J_1(q a) \ \mathcal{F}(k) / (\mathrm{i} k)]
          \ Y_0(q r)
      \end{array} \right\} &
      r > a,
    \end{array}
  \right.                                               \label{eq:solution}
\end{equation}
where $J_n$ and $Y_n$ are the Bessel functions of first and second kind
\cite{HandMathFunc}.
The integration constant $C_1$ is
\begin{eqnarray}
  C_1 & = & 2 \pi^2 n_0 q a \ \chi(q) \ \mathcal{F}(k)/(\mathrm{i} k),
  \quad \mbox{where} \\
  \chi(q) & = & [Y_1(q a) \ J_0(q b) - J_1(q a) \ Y_0(q b)] \ / \ J_0(q b).
\end{eqnarray}

%
%

\subsection{Calculation of the radial force}
The zeroth order calculation of the radial force, $F_s^0$, neglects the 
electron radial motion induced by the space charge and Lorentz force,
i.e. the electron bunch moves as a solid body.
$F_s^0$ is the sum of the radial space charge force and the Lorentz force
arising from the poloidal magnetic field $B_\theta^0$ and the longitudinal
velocity of the electrons:
\begin{equation}
  F_s^0 = e (E_r^0 - \beta B_\theta^0 ),
\end{equation}
where $e$ is the electronic charge with appropriate sign.
With the help of Eqs. (\ref{eq:er0}), (\ref{eq:btheta0}), and
(\ref{eq:solution}) (for $r < a$), we find for the radial force
\begin{eqnarray}
  F_s^0 & = & -(1 - \beta_0^2) \frac{Q e}{a} \int_{-\infty}^\infty \d k
    \e^{\mathrm{i}ks-(k \sigma)^2/2}
    \chi(\mathrm{i} k / \gamma_0) J_1(\mathrm{i} k r / \gamma_0)
    \nonumber \\
    & = & (1 - \beta_0^2) E_r^0.                        \label{eq:radforce}
\end{eqnarray}
It is simple to verify that the integral is always real and negative, thus
the radial force is directed outward and corresponds to a defocusing of the
electron pulse.
This result is a simple generalization of the one obtained in
\cite{Tao,Reassure} for an infinitely long electron beam.

The first order correction to the radial force includes the effect of the
electron radial motion.
Inspection of the transverse phase space plots, $(x^\prime,x)$, where
$x^\prime = \beta_x / \beta_z$, obtained from the numerical calculations
reveals that the electrons acquire a significant radial motion (see Figure
\ref{fig:trph2}).
Let
\begin{equation}
  B_\theta = B_\theta^0 + B_\theta^1,                     \label{eq:btheta}
\end{equation}
where $B_\theta^1$ is the correction term arising from a non-vanishing
radial velocity of the electrons.
$B_\theta^1$ is obtained from Ampere's law
\begin{equation}
  \nabla \times \vec{B} = \frac{4 \pi}{c} \vec{J}
    + \frac{1}{c} \frac{\partial \vec{E}}{\partial t},
\end{equation}
whose projection along the radial direction is
\begin{equation}
  \frac{\partial B_\theta}{\partial z}
  + \frac{1}{c} \frac{\partial E_r}{\partial t}  =
  - \frac{4 \pi}{c} J_r.                                  \label{eq:ampere}
\end{equation}
We substitute Eq. (\ref{eq:btheta}) into (\ref{eq:ampere}) and observe that
the homogenous part of Eq. (\ref{eq:ampere}) is satisfied by $B_\theta^0$,
and the result is
\begin{equation}
  \frac{\partial B_\theta^1}{\partial z} = - \frac{4 \pi \beta}{c} J_r.
                                                         \label{eq:ampere2}
\end{equation}

The radial momentum equation for the electrons is
$\frac{\d}{\d t} (m \gamma v_r) = e E_r (1 - \beta_0)$.
The numerical calculations indicate that the electron radial excursion and
its energy change is quite limited in the initial phases of the pinch.
Hence, the radial momentum equation can be integrated at the onset of the
constriction with the assumption that $\gamma_0$ and $E_r$ are constant
along the electron trajectory.
The resulting $j_r$ is
\begin{equation}
  j_r = - \frac{e}{\gamma_0 m c} \frac{1 - \beta_0}{\beta_0}
    \rho(z,r) E_r^0 z,                                   \label{eq:current}
\end{equation}
where the relationship $\frac{\d}{\d t} = -\beta_0 c \frac{\d}{\d z}$ has
been used.
We substitute Eq. (\ref{eq:current}) into (\ref{eq:ampere2}) and integrate,
and the result, at the center of the electron bunch ($\zeta = 0$), is
\begin{equation}
  B_\theta^1 = \left( \frac{z_c}{z_s} \right)^2 E_r^0,
\end{equation}
where, at $z_c = 0$, the electron bunch is at the beginning of the drift
tube, and
\begin{eqnarray}
  z_s & = &     \left( \frac{\sqrt{2 \pi} I_A a^2 \sigma}{4 Q c}
                \frac{\gamma \beta}{1 - \beta} \right)^{1/2}
                \nonumber \\
      & \cong & \left( \frac{\sqrt{2 \pi} I_A a^2 \sigma}{2 Q c}
                \gamma^3 \beta \right)^{1/2},                 \label{eq:zs}
\end{eqnarray}
where $I_A = m c^3 / e$ is the Alfven current with value $1.7 \times 10^4$
[Amps], or $3.12 \times 10^{13}$ [StatAmps].

We substitute Eqs. (\ref{eq:zs}) and (\ref{eq:btheta}) into the Lorentz
force and obtain the first order correction to Eq. (\ref{eq:radforce}), the
electron radial force:
\begin{equation}
  F_s = \left( 1 - \beta^2
    -\beta \left( \frac{z_c}{z_s} \right)^2 \right) E_r^0.
                                                       \label{eq:radforce2}
\end{equation}
Eq. (\ref{eq:radforce2}) indicates that, at sufficiently large $z_c$
values, the radial force will be turned inward.
The expression for $z_s$ shows that high charge and low energy beams will
become more constricted than their counterparts with lower charge and
higher energies.
For an electron bunch with Gaussian longitudinal charge density
distribution, such as the one considered in this paper, $E_r^0$ peaks at
the center of the pulse. 
Consequently, the maximum constriction will also occur at the center of the
pulse as has been observed in the numerical calculations.

%
%

\subsection{Calculation of the space charge potential}
An estimate of the initial space charge potential at the beginning of the
drift region has been made via the following procedure.
The space charge potential $\Phi^\prime$ is first computed in the electron
rest frame $\mathbf{K}^\prime$ which moves along $z$ with velocity
$v = \beta_0 c$ relative to the laboratory frame $\mathbf{K}$.
In $\mathbf{K}^\prime$, the electrons are stationary, the electron relative
motion is neglected, the electric potential is a well-defined quantity, and
it satisfies the Poisson equation.
$\Phi^\prime$ is then transformed back into the laboratory frame
$\mathbf{K}$, where, from energy conservation, it is translated to an
equivalent $\Delta \gamma$.
The details of the procedure are as follows:
\begin{equation}
  \Phi^\prime(\zeta^\prime,r)
  = - \int_{-\infty}^{\zeta^\prime} \d \eta E_z^\prime (\eta,r)
  = \Phi^\prime(\zeta,r)
  =  - \gamma \int_{-\infty}^\zeta \d \eta E_z (\eta,r),
\end{equation}
where $\Phi^\prime(\zeta,r)$ transforms as the fourth component of a vector
in the four-dimensional space-time space.
Thus, we obtain
\begin{equation}
  \Phi(\zeta,r) =  - \gamma^2 \int_{-\infty}^\zeta \d \eta E_z (\eta,r).
\end{equation}
Energy conservation gives
\begin{equation} 
  \gamma_0 m c^2 = \gamma m c^2
  - e \gamma^2 \int_{-\infty}^\zeta \d \eta E_z (\eta,r).
\end{equation}
$E_z(\eta,r)$ is obtained from Eq. (\ref{eq:solution})
(for $r < a$).
Thus,
\begin{eqnarray}
  \Delta \gamma(\zeta,r) & = & \frac{2 Q e \gamma_0}{m c^2 a}
  \int_{-\infty}^\infty \d k
  \left( \e^{\mathrm{i}k\zeta}- \e^{-\mathrm{i}k\zeta_0} \right)
    / (\mathrm{i}k) \nonumber \\
  & \times &
  \left[ \pi \chi(q) J_0(q r) + 2 \gamma / (\mathrm{i} a k) \right],
\end{eqnarray}
where $\zeta_0$ is arbitrary, provided it satisfies the condition
$\zeta_0 >> \sigma$.
Finally, we compute the normalized momentum $z$-component
\begin{equation}
  \gamma \beta_z = \left[ \left( \gamma_0
  + \Delta \gamma(\zeta,r) \right)^2 -1 \right]^{1/2}.      \label{eq:zmom}
\end{equation}

In the derivation of Eq. (\ref{eq:zmom}), the electrons are assumed to be
strictly stationary in $\mathbf{K}^\prime$.
When significant radial motion is present, the results of this analysis
will be inaccurate as is observed in the numerical results.

%
%

\section{Numerical simulation results}
\label{sec:numerics}
The numerical calculations of a cylindrical, axisymmetric electron bunch in
a drift tube with perfect wall conductivity were performed with the
``particle in a cell'' code ARGUS \cite{ARGUS}.
ARGUS is a fully three-dimensional and time-dependent $(x,y,z,t)$ solver
for Maxwell's equations.
We used the ``pic'' solver option.
  
We also compared ARGUS results to the calculations from the standard code
PARMELA \cite{PARMELA}.
PARMELA was originally developed at Los Alamos National Laboratory and is
widely used within the accelerator community.
In PARMELA, the forces between the electrons are computed in the electron
rest frame where all the electron relative motion is neglected.

%
%

\subsection{ARGUS results}
The electron bunch is generated at $z = 0$, and the electron density
conforms to Eq. (\ref{eq:density}).
The total electron bunch length in our simulation is 5.0 $\sigma$'s.
$\sigma$ varies from $0.1274$ to $0.7644$ [cm].
The bunch radius $a$ varies from $0.1$ to $0.3$ [cm], the electron energy
$E$ from $20.$ to $100.$ [MeV], and the charge $Q$ from $2.0$ to $18.0$
[nC].
The electrons are injected with uniform $\beta_z$, and
$\beta_x = \beta_y = 0$.

\begin{figure}[h]
\begin{center}
\epsfig{file=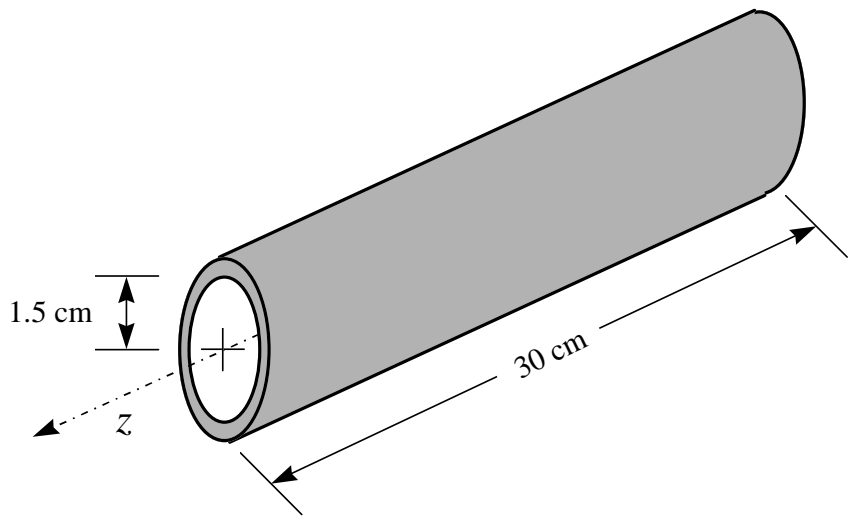,width=4.65in,bb=139 145 657 466,clip=true}
\end{center}
\caption{Drift tube geometry in ARGUS. The positive $z$-axis is oriented in
the direction of the bunch motion.} \label{fig:tube}
\end{figure}

In the code the electrons are represented by ``macro particles''.
The density of the macro particles is uniform, but the charge is weighted
to reproduce the appropriate charge density distribution.
The drift tube, as represented in ARGUS, is shown in Figure \ref{fig:tube}.

%
%

The drift tube radius is $1.5$ [cm] and the length is $30.0$ [cm].
The number of grid points along $x$ and $y$ is 57, and the number along $z$
varies from $192$ to $571$.
The spacing in $x$ and $y$ is not uniform:
the smaller spacing ($0.02$ [cm]) is used in proximity of the $z$-axis of
the tube, and the larger spacing ($0.163$ [cm]) in the vicinity of the
wall.
The spacing along $z$ is constant.
The time step, $0.35$ [ps], is selected such that the Courant condition is
satisfied everywhere.
This condition is $v_p \Delta t \le \Delta x_\mathrm{min} / 2$, where $v_p$
is the particle velocity, $\Delta t$ is the time step, and
$\Delta x_\mathrm{min}$ is the smallest grid spacing.

\begin{figure}[ht]
\begin{center}
\epsfig{file=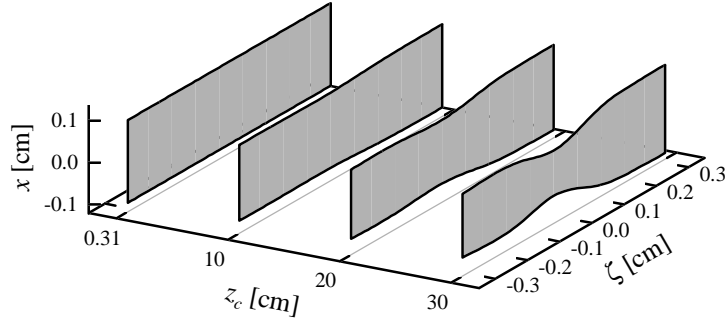,width=4.0in,bb=76 150 652 416,clip=true}
\end{center}
\caption{Evolution of the bunch envelope.  Bunch parameters: $Q = 2.0$
[nC], $a = 0.1$ [cm], $\sigma = 0.1274$ [cm], $b = 1.5$ [cm], $E = 20.$
[MeV], $\gamma_0 = 40.138$ .}
\label{fig:bunchenv}
\end{figure}

In Figure \ref{fig:bunchenv} the evolution of the $(x,\zeta)$ bunch
envelope at different positions in the drift is shown.
In view of the cylindrical symmetry of the electron bunch, the $(x,\zeta)$
envelope is identical to the $(y,\zeta)$ one.
At $z_c = 0.$ the cross section is uniform, in the interval
$10. < z_c < 20.$ [cm] a small pinch develops, and at $z_c = 30.$ [cm] a
fully evolved pinch has been established.
The largest pinch occurs in the middle of the bunch in correspondence to
the maximum radial force as also seen in Eq. (\ref{eq:radforce2}).
A close inspection of the bunch envelope reveals a tapering of the rear
half.
The wake fields from the front half may be responsible for this behavior.

%
%

\begin{figure}[h]
\begin{center}
\epsfig{file=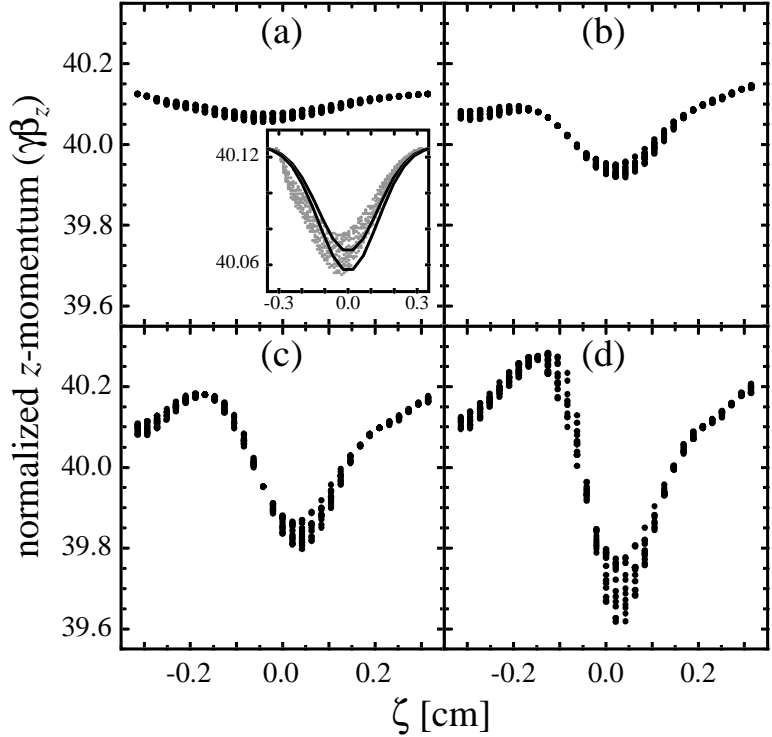,width=4.2in,bb=108 63 612 549,clip=true}
\end{center}
\caption{Evolution of the longitudinal phase space at drift positions (a)
$z_c = 0$, (b) $z_c = 10.$ [cm], (c) $z_c = 20.$ [cm], and (d) $z_c = 30.$
[cm].  Bunch parameters: $Q = 2.0$ [nC], $a = 0.1$ [cm], $\sigma = 0.1274$
[cm], $b = 1.5$ [cm], $E = 20.$ [MeV], $\gamma_0 = 40.138$ .}
\label{fig:longph}
\end{figure}

In Figure \ref{fig:longph} the evolution of the longitudinal phase space,
$\gamma \beta_z$ vs $\zeta$, is presented for the same case as in Figure
\ref{fig:bunchenv}.
In the inset of Figure \ref{fig:longph}(a) we have overlapped, with an
expanded scale, the results of Eq. (\ref{eq:zmom}) with the ARGUS
calculations for purpose of comparison.
Here, the lower curve corresponds to $r = 0$ and the upper one to $r = a$.
The agreement is quite satisfactory in view of the simplifying assumptions
made in our analytical formulation.
As the pulse drifts in the tube, the relative electron motion increases,
and a more complex longitudinal phase space evolves.

%
%

\begin{figure}[h]
\begin{center}
\epsfig{file=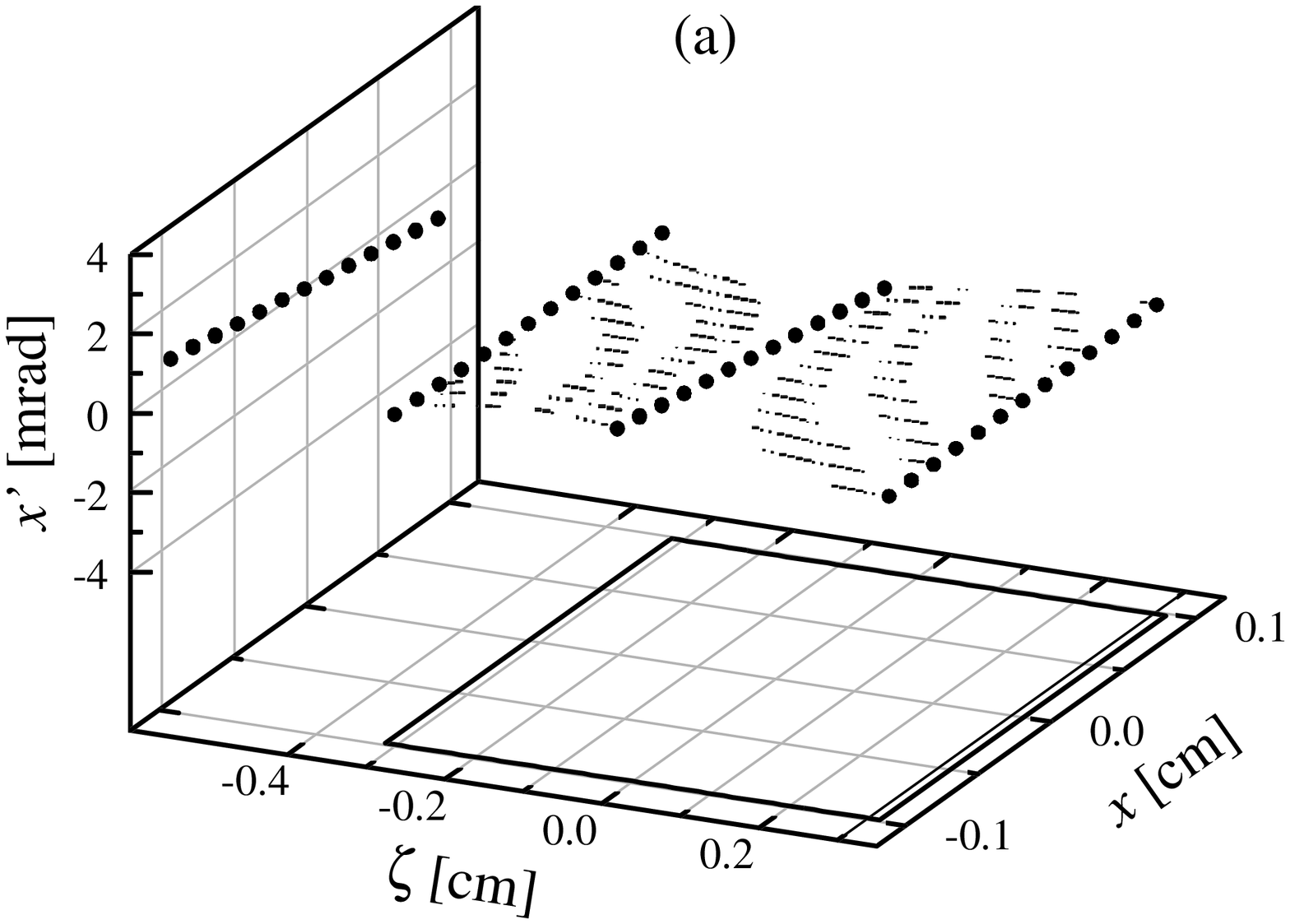,width=2.5in,bb=90 93 657 499,clip=true}
\epsfig{file=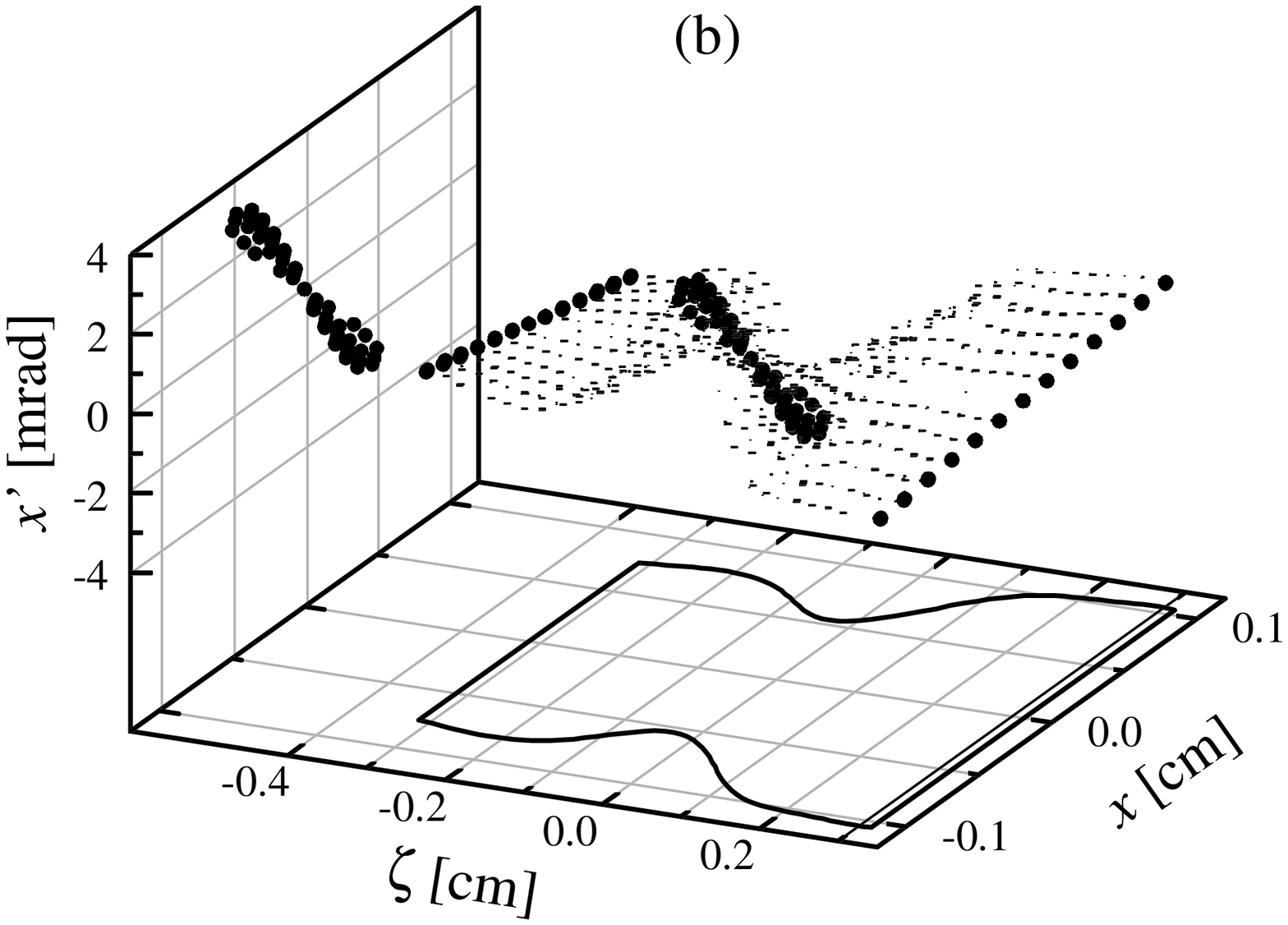,width=2.5in,bb=90 93 657 499,clip=true}
\end{center}
\caption{Evolution of the $(x^\prime,x,\zeta)$ phase space at drift
positions (a) $z_c = 0$ and (b) $z_c = 30.$ [cm].  The pulse envelope is
projected on the bottom panel.  Bunch parameters: $Q = 2.0$ [nC], $a = 0.1$
[cm], $\sigma = 0.1274$ [cm], $b = 1.5$ [cm], $E = 20.$ [MeV],
$\gamma_0 = 40.138$ .}
\label{fig:trph}
\end{figure}

In Figure \ref{fig:trph} the phase space $(x^\prime,x,z)$ is shown at
$z_c = 0$ and $z_c = 30.$ [cm].
At $z_c = 0$ the phase space is planar as expected from the initial
conditions inposed on the momemtum of the electrons.
At $z_c = 30.$ [cm] the phase space has become twisted.
A projection of the intersection of the phase space with the $\zeta = 0$
plane, where the largest pinch occurs, is displayed on the rear panel of
the figure.
A significant inward motion of the electrons is present.

%
%

\begin{figure}[h]
\begin{center}
\epsfig{file=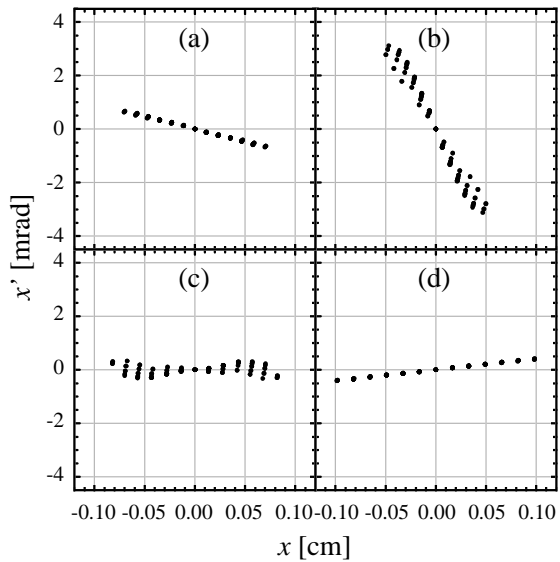,width=3.1in,bb=121 68 620 559,clip=true}
\end{center}
\caption{Intersections of the $(x^\prime,x,\zeta)$ phase space at drift
position $z_c = 30.$ [cm].  Intersections are at (a) $z = -0.315$ [cm], (b)
$z = -0.030$ [cm], (c) $z = +0.094$ [cm], and (d) $z = +0.315$ [cm].  Bunch
parameters: $Q = 2.0$ [nC], $a = 0.1$ [cm], $\sigma = 0.1274$ [cm],
$b = 1.5$ [cm], $E = 20.$ [MeV], $\gamma_0 = 40.138$ .}
\label{fig:trph2}
\end{figure}

In Figure \ref{fig:trph2} the intersections of the phase space
$(x^\prime, x, z)$ with planes at constant $\zeta$ are shown when the pulse
is at the end of the drift tube, $z_c = 30.$ [cm].
Figure \ref{fig:trph2}(a) and \ref{fig:trph2}(b) correspond to the tail and
the middle of the bunch, respectively.
The entire $(x^\prime,x)$ phase space is shown in Figure
\ref{fig:trphcomp}(a).

%
%

It is convenient to introduce the concept of the fractional rms ``slice''
emittance.
The total normalized rms emittance of the pulse is defined as
\begin{equation}
  \varepsilon_\mathrm{rms}^N = \gamma \beta
  \left[ \pi \left( \left< x^2 \right> \left<x^{\prime 2} \right>
    - \left< x x^\prime \right>^2 \right)^{1/2} \right],
\end{equation}
where the averages indicated by the angle brackets, $\left< \right>$,
extend over all the pulse particles.
The fractional slice emittance is defined as
\begin{equation}
  \varepsilon_{\mathrm{rms},f}^N (\zeta,\Delta \zeta)
  = \gamma \beta
  \left[ \pi \left( \left< x^2 \right>_{\Delta \zeta}
                    \left< x^{\prime 2} \right>_{\Delta \zeta}
                - \left< x x^\prime \right>^2_{\Delta \zeta} \right)^{1/2}
  \right] / \varepsilon_\mathrm{erms}^N,
\end{equation}
where the averages indicated by the angle brackets,
$\left< \right>_{\Delta \zeta}$, are limited to particles within the slice,
i.e. with $\zeta$-coordinate between $\zeta$ and $\zeta + \Delta \zeta$.

\begin{figure}[h]
\begin{center}
\epsfig{file=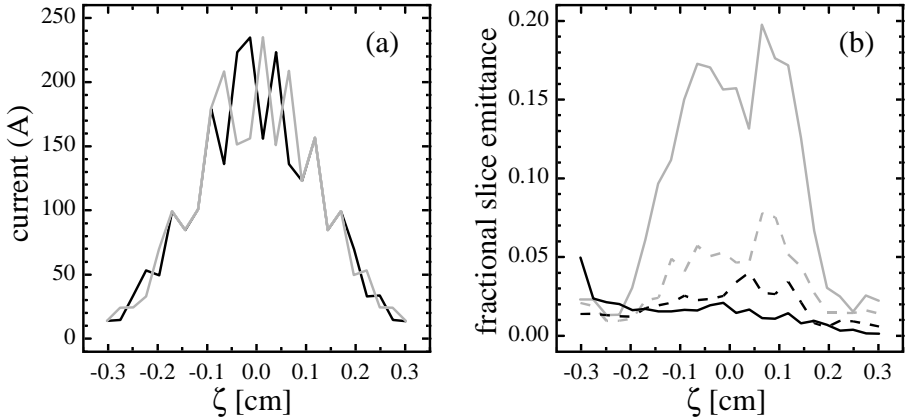,width=4.85in,bb=37 123 764 472,clip=true}
\end{center}
\caption{(a) Bunch current and (b) fractional slice emittance vs position
in the bunch at different locations in the drift tube.  All fractional
slice emittances in (b) are normalized to $18.5$ [$\pi$ mm-mrad].  Legend:
dark solid line, $z_c = 0.66$ [cm]; dark broken line, $z_c = 10.$ [cm];
light broken line, $z_c = 20.$ [cm]; light solid line, $zc = 30.$ [cm].
Bunch parameters: $Q = 2.0$ [nC], $a = 0.1$ [cm], $\sigma = 0.1274$ [cm],
$b = 1.5$ [cm], $E = 20.$ [MeV], $\gamma_0 = 40.138$ .  Curve fluctuations
are due to discretization noise.}
\label{fig:slices}
\end{figure}

In Figure \ref{fig:slices}(a) the pulse current profile at $z_c = 0$ and
$z_c = 30.$ [cm] is shown.
The current profiles are nearly identical, as expected, because of the
limited relative longitudinal motion among the particles.
In Figure \ref{fig:slices}(b) the fractional slice emittance as a function
of $\zeta$ for different position $z_c$ along the drift tube is shown.
Observe that the fractional slice emittance is always significantly smaller
than the total pulse emittance of $18.5$ [$\pi$ mm-mrad] at the end of the
drift tube.
This indicates that the pulse emittance is strongly correlated.
 
%
%

\begin{figure}[ht]
\begin{center}
\epsfig{file=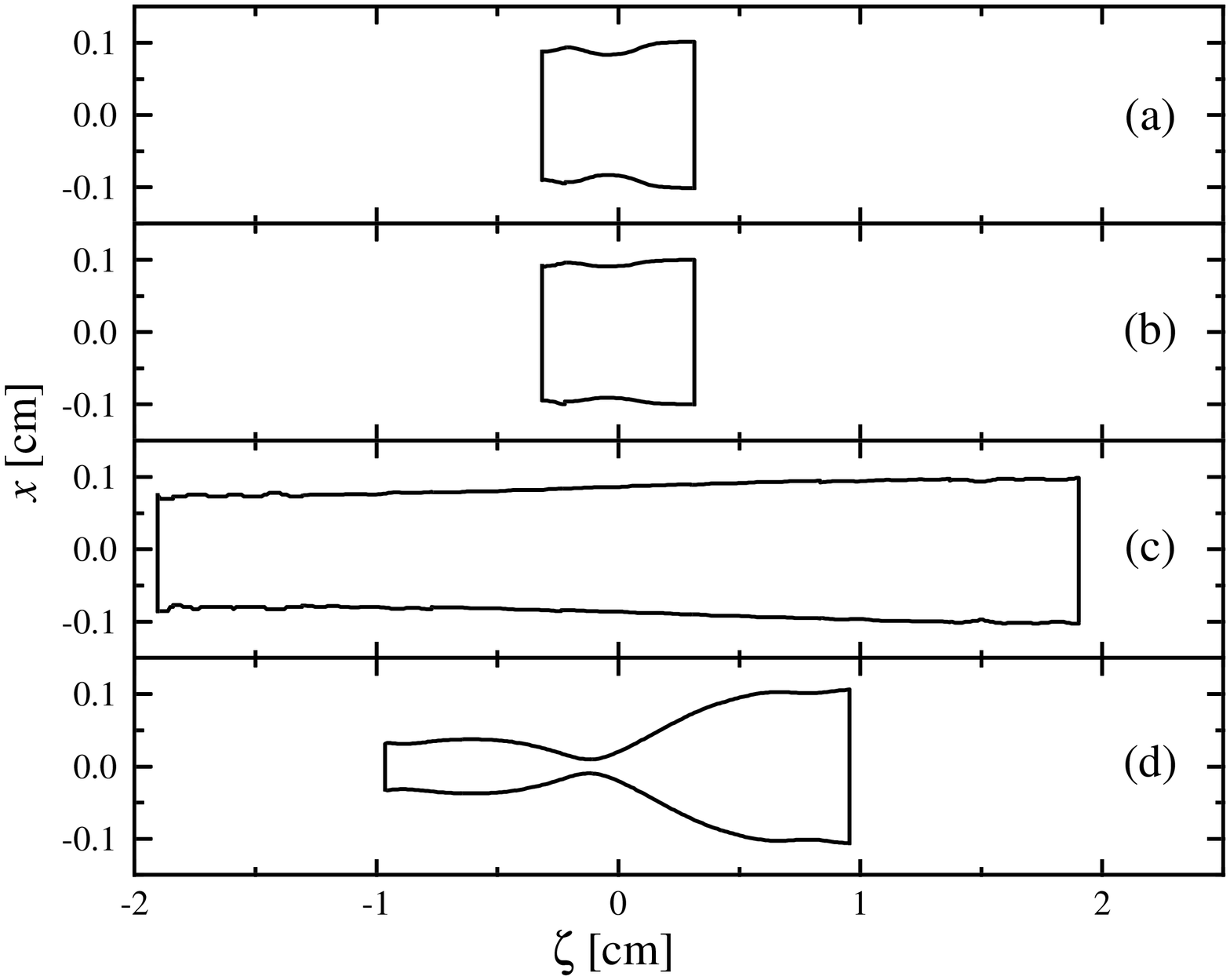,width=3.8in,bb=61 45 722 570,clip=true}
\end{center}
\caption{Various longitudinal $(\zeta,x)$ bunch envelopes at drift location
$z_c = 30.$ [cm].  For (a)--(d), $a = 0.1$ [cm].  Bunch parameters:
\protect\\ (a) $Q = 2.0$ [nC], $\sigma = 0.1274$ [cm], $b = 0.25$ [cm], $E
= 50.$ [MeV], $\gamma_0 = 98.84$ \protect\\ (b) $Q = 2.0$ [nC], $\sigma =
0.1274$ [cm], $b = 0.25$ [cm], $E = 100.$ [MeV], $\gamma_0 = 196.69$
\protect\\ (c)  $Q = 2.0$ [nC], $\sigma = 0.7639$ [cm], $b = 0.25$ [cm], $E
= 20.$ [MeV], $\gamma_0 = 40.13$ \protect\\ (d) $Q = 6.0$ [nC], $\sigma =
0.3822$ [cm], $b = 1.5$ [cm], $E = 20.$ [MeV], $\gamma_0 = 40.13.$}
\label{fig:bunchenv2}
\end{figure}

In Figure \ref{fig:bunchenv2} typical longitudinal $(\zeta,x)$ bunch
envelopes at $zc = 30.$ [cm] are presented, corresponding to varying
energies, bunch lengths, and charges.
The pinch decreases at higher energies and increases with bunch length for
constant charge density.

%
%

\begin{figure}[h]
\begin{center}
\epsfig{file=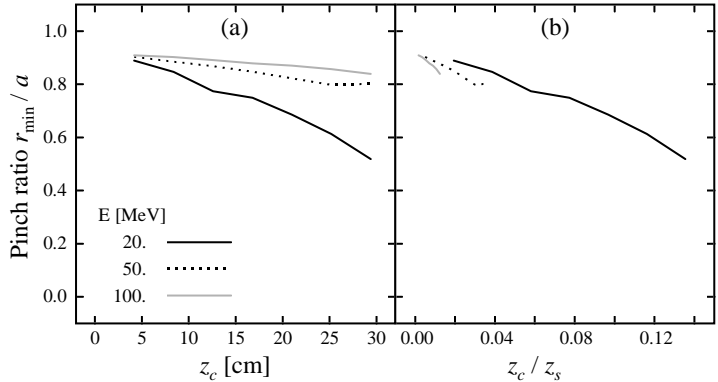,width=3.85in,bb=50 106 752 495,clip=true}
\end{center}
\caption{Pinch ratio vs (a) drift position $z_c$ and (b) normalized drift
position $z_c/z_s$ in the drift tube for various initial energies.  Common
bunch parameters: $Q = 2.0$ [nC], $a = 0.1$ [cm], $\sigma = 0.1274$ [cm],
$b = 0.25$ [cm].}
\label{fig:pinch1}
\end{figure}

The evolution of the pinch ratio as a function of the position in the drift
tube is shown in Figure \ref{fig:pinch1}(a).
Here, $r_\mathrm{min} / a$, the ratio of the minimum bunch envelope radius
divided by the initial bunch radius, is plotted versus the position in the
drift tube for three different values of the bunch energy.
Higher energy bunches undergo less pinching than lower energy ones.
In Figure \ref{fig:pinch1}(b), $r_\mathrm{min} / a$ is plotted versus the
normalized bunch center position, $z_c / z_s$.
The three separate curves of Figure \ref{fig:pinch1}(a) nearly collapse
into a single one, thus showing that Eq. (\ref{eq:zs}) correctly captures
the energy scaling of the pinch ratio.

%
%

\begin{figure}[h]
\begin{center}
\epsfig{file=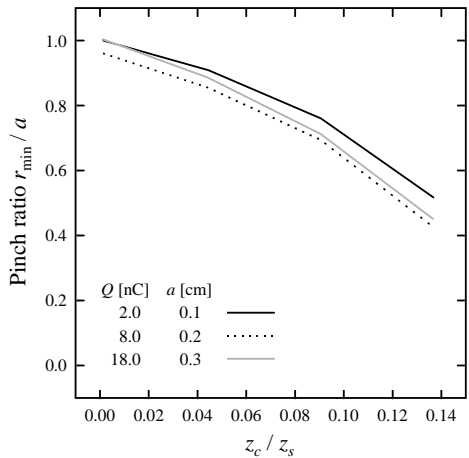,width=2.55in,bb=127 44 634 541,clip=true}
\end{center}
\caption{Pinch ratio vs normalized drift position zc/zs in the drift tube
for various charges and initial bunch radii.  Common bunch parameters:
$\sigma = 0.1274$ [cm], $b = 1.5$ [cm], $E = 20.$ [MeV], $\gamma_0 =
40.138$.}
\label{fig:pinch2}
\end{figure}

In Figure \ref{fig:pinch2} the evolution of the pinch ratio is shown as a
function of $z_c / z_s$ for pulses of different diameters, same lengths and
different charges selected to maintain a constant electron density.
As in the previous case, the pinch ratio is well parameterized by
$z_c  / z_s$.

%
%

\begin{figure}[h]
\begin{center}
\epsfig{file=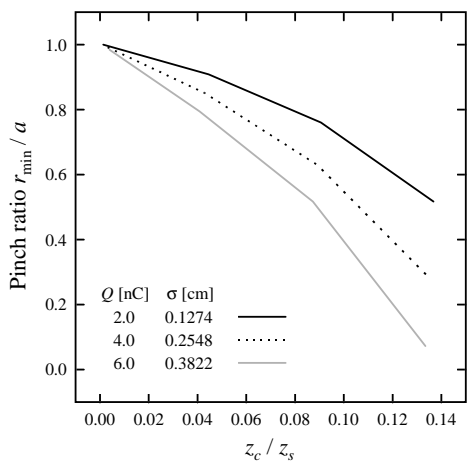,width=2.55in,bb=127 44 634 541,clip=true}
\end{center}
\caption{Pinch ratio vs position $z_c / z_s$ in the drift tube for various
charges and bunch lengths.  Common bunch parameters: $a = 0.1$ [cm], $b =
1.5$ [cm], $E = 20.$ [MeV], $\gamma_0 = 40.138$.}
\label{fig:pinch3}
\end{figure}

Finally, in Figure \ref{fig:pinch3} the evolution of the pinch ratio is
shown as a function of $z_c / z_s$ for bunches of different lengths,
identical transverse cross sections and different charges selected to
maintain a constant electron density.
Unlike the previous cases, the pinch ratio is not well parameterized by
$z_c / z_s$.
The probable explanation for the inability to parameterize the present case
is that the wake fields of the particles in the leading edge of the bunch
are not properly accounted for in our simple theory.

%
%
%
%
%

%
%

\subsection{Comparison of ARGUS and PARMELA results}
\begin{figure}[h]
\begin{center}
\epsfig{file=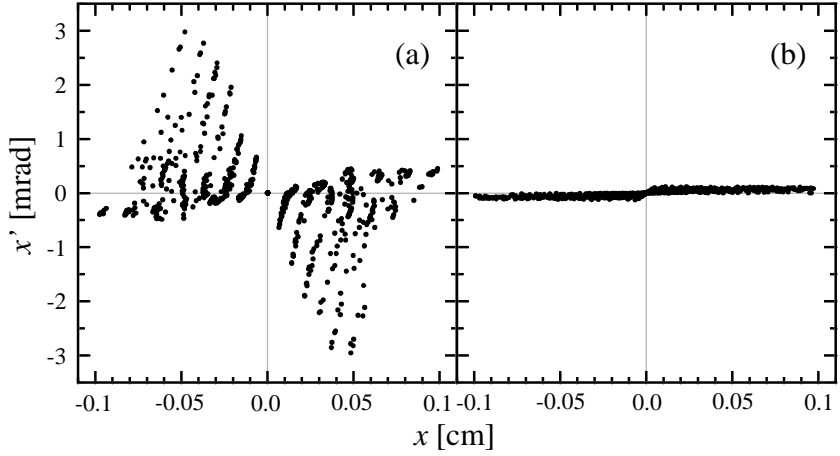,width=4.55in,bb=37 120 699 489,clip=true}
\end{center}
\caption{(a) ARGUS and (b) PARMELA results of transverse $(x,x^\prime)$
phase space at drift position $z_c = 30.$ [cm].  Bunch parameters: $Q =
2.0$ [nC], $a = 0.1$ [cm], $\sigma = 0.1274$ [cm], $b = 1.5$ [cm], $E =
20.$ [MeV], $\gamma_0 = 40.138$.}
\label{fig:trphcomp}
\end{figure}

ARGUS and PARMELA yield quite different results for identical bunch
conditions.
In Figure \ref{fig:trphcomp} the transverse phase space, as computed by
ARGUS and PARMELA at $z_c = 30.$ [cm], is shown.
The PARMELA phase space reflects only a mild defocusing effect due to the
space charge.
The ARGUS counterpart combines the focusing and defocusing due to the
pinching force.

%
%

\begin{figure}[h]
\begin{center}
\epsfig{file=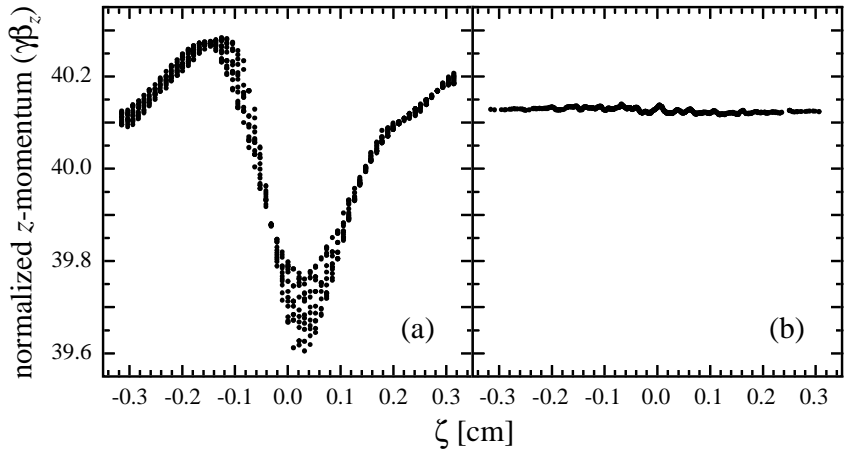,width=4.55in,bb=51 111 730 484,clip=true}
\end{center}
\caption{(a) ARGUS and  (b) PARMELA results of longitudinal $(\gamma
\beta_z,\zeta)$ phase space at drift position $z_c = 30.$ [cm].  Bunch
parameters: $Q = 2.0$ [nC], $a = 0.1$ [cm], $\sigma = 0.1274$ [cm], $b =
1.5$ [cm], $E = 20.$ [MeV], $\gamma_0 = 40.138$.}
\label{fig:longphcomp}
\end{figure}

In Figure \ref{fig:longphcomp} the longitudinal phase space, as computed by
ARGUS and PARMELA at $z_c = 30.$ [cm], is shown.
Here, PARMELA cannot reproduce the complexity of the longitudinal momentum
distribution within the bunch.

%
%

\begin{figure}[h]
\begin{center}
\epsfig{file=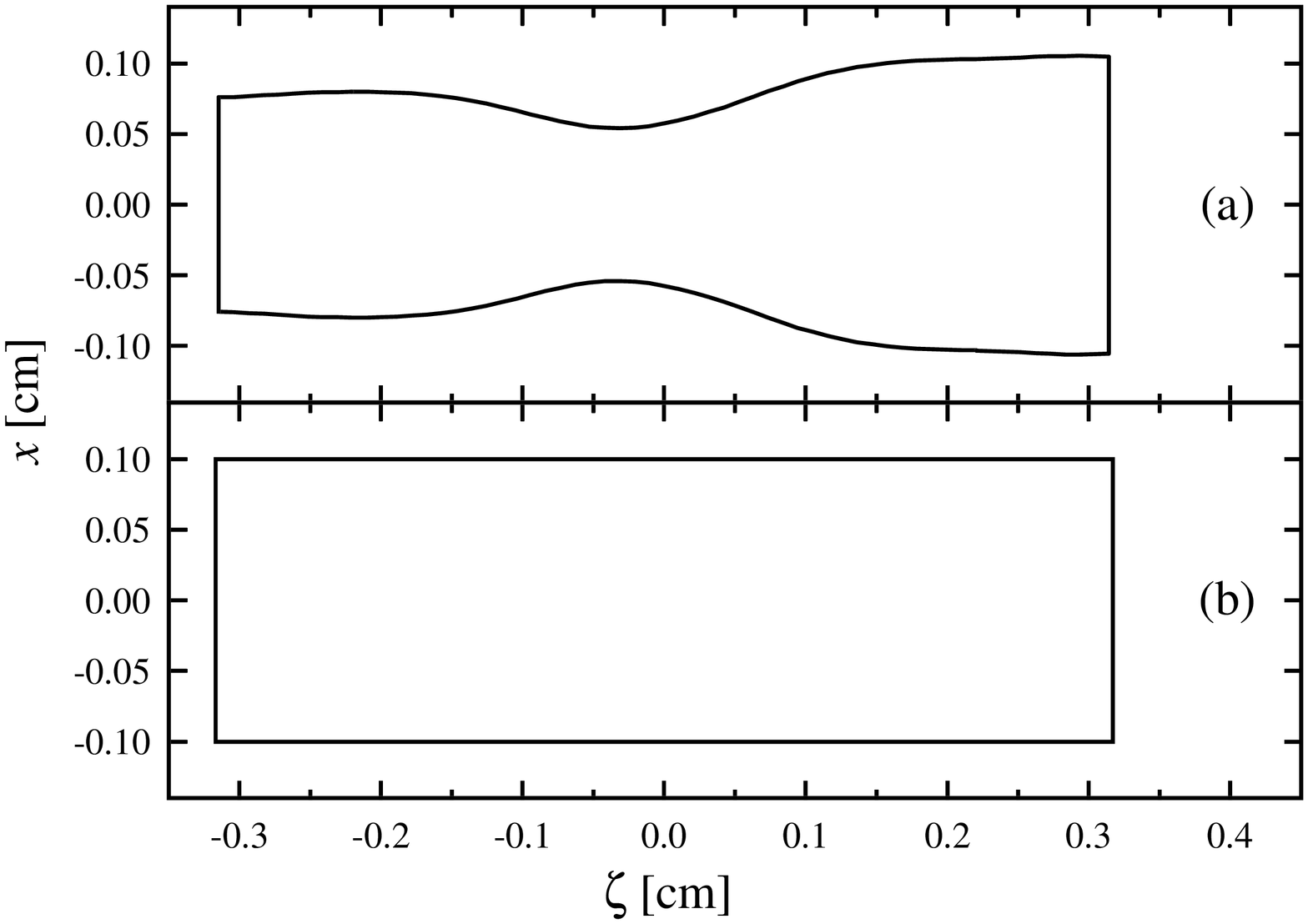,width=4.6in,bb=79 85 741 556,clip=true}
\end{center}
\caption{(a) ARGUS and (b) PARMELA results of longitudinal $(\zeta ,x)$
bunch envelopes at drfit position $z_c = 30.$ [cm].  Bunch parameters: $Q =
2.0$ [nC], $a = 0.1$ [cm], $\sigma = 0.1274$ [cm], $b = 1.5$ [cm], $E =
20.$ [MeV], $\gamma_0 = 40.138$.}
\label{fig:envcomp}
\end{figure}

Finally, in Figure \ref{fig:envcomp} the longitudinal $(\zeta,x)$ envelopes
at $z_c = 30.$ [cm], as computed by ARGUS and PARMELA, are presented.
No pinching of the bunch is exhibited by the PARMELA calculations.

%
%
%
%
%

%
%

\section{Summary and Conclusions}
\label{sec:summary}
We have presented the results of extensive numerical simulations of
electron bunches with charge densities $\rho$ of the order of $10^2$ to
$10^3$ [nC/cm$^3$] and energies between $20.$ and $100.$ [MeV].
The results indicate the presence of a strong pinch in the middle of an
electron bunch with a Gaussian longitudinal electron density distribution.
The pinching force scales approximately as $\gamma^{-3/2}$ and $r^{1/2}$.
This force generates an increase in the correlated bunch emittance, and the
space charge depression is also affected.
A simplified analysis to explain these results has been described.
The results are in significant disagreement with the ones obtained from the
standard accelerator code PARMELA.
This is attributed to the neglect of the electron relative motion in the
space charge calculations performed in PARMELA.

The pinch effects we have uncovered will play a significant role in the
photoinjectors used in Free Electron Laser and other high-brightness
accelerators.
Here, the electron pulse has very low energy, and this will limit the
allowed charge density in order to avoid undue pinching and space charge
effects.
The break up of the electron bunch at the photoinjector has been observed
\cite{Loulergue,Dowell} and attributed solely to the longitudinal component
of the space charge force.
The pinching force discussed in this paper may also play a role.
In addition, the design of magnetic pulse compressors will need to account
for the pinch which may become significant in the last stage of the
compression.

Further analytical studies and numerical simulations are necessary to
better understand (a) the evolution of the pinch as $z_c \rightarrow
\infty$ and (b) the effect of the longitudinal electron density
distributions.
Preliminary indications are that uniform longitudinal charge density
distributions become pinched at multiple positions in the bunch.
The pinch period depends upon the pulse parameters.
More importantly, experimental observation of the pinch is needed to
completely validate these results.


\begin{ack}
We wish to thank Robert Snead for supplying computer time at the USA-SSDC
facilities, Dr. John Petillo for technical support for ARGUS, and Dr. Art
Vetter and Dr. Dave Dowell for helpful discussions.
This work was partially performed under contract number DASG-60-90-c-0106.
\end{ack}



\end{document}